\documentclass[pdftex,twocolumn,showpacs, showkeys,epjc3]{svjour3}
\RequirePackage[T1]{fontenc}

\smartqed
\usepackage[T1]{fontenc}
\usepackage[latin1]{inputenc}
\usepackage{graphicx}
\usepackage[english]{babel}
\usepackage{bm}
\usepackage{amsmath}
\usepackage{amssymb}
\usepackage{amsfonts}
\usepackage{epsfig}
\usepackage{colordvi}
\usepackage{color}

\RequirePackage{graphicx}
\RequirePackage{mathptmx}      
\RequirePackage{flushend}
\RequirePackage[numbers,sort&compress]{natbib}
\RequirePackage[colorlinks,citecolor=blue,urlcolor=blue,linkcolor=blue]{hyperref}

\begin{document}

\title{Noether Symmetries in Gauss-Bonnet-teleparallel cosmology}


\author{
Salvatore Capozziello\thanksref{e1,addr1,addr2,addr3,addr4},\,Mariafelicia De Laurentis\thanksref{e2,addr2,addr4,addr5,addr6},\,Konstantinos F. Dialektopoulos \thanksref{e3,addr1,addr2}
}

\thankstext{e1}{e-mail: capozziello@na.infn.it}
\thankstext{e2}{e-mail: laurentis@th.physik.uni-frankfurt.de}
\thankstext{e3}{e-mail: dialektopoulos@na.infn.it}

\institute{Dipartimento di Fisica "E. Pancini", Universita'
		   di Napoli {}``Federico II'', Compl. Univ. di
		   Monte S. Angelo, Edificio G, Via Cinthia, I-80126, 			   Napoli, Italy.\label{addr1}
          \and
		  INFN Sezione  di Napoli, Compl. Univ. di
		   Monte S. Angelo, Edificio G, Via Cinthia, I-80126, 			   Napoli, Italy.\label{addr2}
          \and
		  Gran Sasso Science Institute (INFN), Via F. Crispi 			   	   7, I-67100, L' Aquila, Italy.\label{addr3}
		  \and
		  Tomsk State Pedagogical University, 634061 Tomsk, 				   Russia.\label{addr4}
		   \and		  
		   Institute for Theoretical Physics,
		   Goethe University, Max-von-Laue-Str.~1, D-60438 				   Frankfurt, Germany.\label{addr5}
		   \and
		   Lab.Theor.Cosmology,Tomsk State University of Control 		   Systems and Radioelectronics (TUSUR), 634050 Tomsk, 			   Russia.\label{addr6}
}

\date{Received: date / Accepted: date}

\maketitle

\begin{abstract}
  A generalized teleparallel cosmological model, \\$f(T_\mathcal{G},T)$, containing the torsion scalar $T$ and the teleparallel counterpart of the Gauss-Bonnet topological invariant $T_{\mathcal{G}}$, is studied in the framework of the Noether Symmetry Approach. As  $f(\mathcal{G}, R)$ gravity, where $\mathcal{G}$ is the Gauss-Bonnet topological invariant and $R$ is the Ricci curvature scalar,  exhausts all the   curvature information that one can construct from the Riemann tensor,  in the same  way, $f(T_\mathcal{G},T)$ contains all the  possible information directly related to the torsion tensor. In this paper, we  discuss how  the Noether Symmetry Approach allows to fix the form of the function  \\ $f(T_\mathcal{G},T)$ and to derive exact cosmological solutions.   
\end{abstract}


\section{Introduction}
\label{uno}
Extended theories of gravity are  semi-classical  approaches   where the effective gravitational Lagrangian is modified, with respect to the Hilbert-Einstein one, by considering higher order terms of  curvature invariants, torsion tensor, derivatives of curvature invariants and scalar fields (see for example \cite{RepProgPhys,PhysRepnostro, OdintsovPR, annalen}). In particular, taking into account the Ricci, Riemann and Weyl invariants, one can construct terms 
 like $R^2,$ $R^{\mu\nu}R_{\mu\nu},$ $R^{\mu\nu\delta\sigma}R_{\mu\nu\delta\sigma}$, $W^{\mu\nu\delta\sigma}W_{\mu\nu\delta\sigma},$ that give rise to fourth-order theories in the metric formalism  \cite{stelle,arturo}. Considering   minimally or nonminimally  coupled scalar fields to the geometry, we deal with scalar-tensor theories of gravity \cite{faraoni,maeda}. Considering terms like    $R\Box R,$ $R\Box^k R$, we are dealing with higher-than fourth order theories \cite{schmidt,lambiase}.   $f(R)$ gravity is the simplest class of these models  where a  generic function  of the Ricci scalar $R$ is considered. The interest for these extended models  is related both to the problem of quantum gravity \cite{PhysRepnostro} and to 
the possibility to explain the accelerated expansion of the universe, as well as the structure formation, without invoking new particles  in the matter/energy content of the universe \cite{annalen,stelle,arturo,faraoni,maeda,schmidt,lambiase,cianci,greci,supernovae1,supernovae2,quintessence}. In other words, the attempt is to address the dark side of the universe by changing the geometric sector and remaining unaltered the matter sources with respect to the Standard Model of Particles. However, in the framework of this "geometric picture", the debate is very broad involving the  fundamental structures of gravitational interaction. Just to summarize some points, gravity could be described only by metric (in this case we deal with a metric approach), or by metric and connections (in this case, we are considering a metric-affine approach \cite{francaviglia}), or by a purely affine approach \cite{purely}. Furthermore, dynamics could be related to curvature tensor,  as in the original Einstein theory, to both curvature and torsion \cite{hehl}, or to torsion only, as in the so called {\it teleparallel gravity} \cite{teleparallel}.

Starting from these original theories and motivations, one can build more complex Lagrangians, by using different combinations of curvature scalars and their derivatives, or topological invariants, such us the Gauss-Bonnet term, $\mathcal{G}$, as well as the torsion scalar $T$. Many theories have been proposed  considering generic functions of such terms, like  $f({\cal G})$,  $f(T)$, $f(R, {\cal G})$ and $f(R, T)$ \cite{FGRgravity,17,18,  21,antonio,felixquint, double,ft,Kofinas:2014owa,fRT,sebastian,cimento,hamilton,pla2,marek,  Capozziello:2009te,  2011GReGr..43.2807S, stabile1, zerb}.
However, the problem is {\it how many} and {\it what kind} of geometric invariants can be used, and furthermore what kind of physical information one can derive from them. For example, it is well known that $f(R)$ gravity is the straightforward extension of the Hilbert-Einstein which is $f(R)=R$ and $f(T)$ is the extension of teleparallel gravity which is $f(T)=T$. However, if one wants to consider the whole  information contained in curvature invariants, one has to take into account also combinations of Riemann, Ricci and Weyl  tensors\footnote{Clearly, this means that we are not considering higher-order derivative terms like $\Box R$, or derivative combinations of curvature invariants.}.  As discussed in \cite{double}, assuming a $f(R, {\cal G}$ theory means to consider the whole curvature budget and then all the degrees of freedom related to curvature. 

Assuming the teleparallel formalism, a $f(T_{\cal G},T)$ theory, where $T_{\cal G}$ is the torsional counterpart of the Gauss-Bonnet topological invariant, means to exhaust all the degrees of freedom related to torsion and then completely extend $f(T)$ gravity.
It is important to stress, as we will show below, that the Gauss-Bonnet invariant derived from curvature differs from the same topological invariant derived from torsion in less than a total derivative and then the dynamical information is the same in both representations. According to this result, the topological invariant allows a regularization of dynamics also in  the teleparallel torsion picture (see \cite{double, PhysRevD.28.1876} for a discussion in the curvature representation).

The layout of the paper is the following. In Sec.\ref{tre}, we sketch the basic ingredients of the $f(T_{\cal G},T)$ theory showing, in particular, the equivalence between $T_{\cal G}$ and $\cal G$.   Sec.\ref{noether} is devoted to derive the cosmological counterpart of the theory and to the derivation of the Noether symmetry. The specific forms of   $f(T_{\cal G},T)$ function, selected by the Noether symmetry, are discussed in Sec. \ref{selection}.  Cosmological solutions are given in Sec. \ref{solutions}. Conclusions are drawn in Sec.\ref{conclusion}.

\section{$f(T_\mathcal{G},T)$  gravity }
\label{tre}

In order to incorporate spin in a geometric description, as well as to bring gravity closer to its gauge formulation, people started, some years ago, to study torsion in gravity \cite{teleparallel,hehl}. An extensive review of torsional theories (teleparallel, Einstein-Cartan, metric-affine, etc) is presented in \cite{RepProgPhys}. If in the action of the teleparallel theory, i.e. in a curvature-free \textit{vierbein} formulation, we replace the torsion scalar, $T$, with a generic function of it, we obtain the so called $f(T)$ gravity \cite{16a,Kazuharu:2010b,Dent:2010, Li:2010},

In this paper,  we will study a  theory  whose Lagrangian is a generic function of the Gauss-Bonnet teleparallel term, 
$T_\mathcal{G}$ and the torsion scalar, $T$, i.e.
\begin{equation}\label{action1}
 {\cal A}=\frac{1}{2\kappa}\int d^4 x\sqrt{-g}\left[f(T_\mathcal{G},T)+{\cal L}_m\right]\,,
 \end{equation}
which is a straightforward generalization of 
\begin{equation}
 {\cal A}=\frac{1}{2\kappa}\int d^4 x\sqrt{-g}\left[f(T)+{\cal L}_m\right]\,,
 \end{equation}
 where ${\cal L}_m$ is the standard matter that, in the following considerations,  we will discard.
 It is important to note that the field equations of $f(T)$ gravity are of second order in the metric derivatives and thus simpler than those of $f(R)$ gravity, which are of fourth order \cite{RepProgPhys}. 
 
 The metric determinant $\sqrt{-g}$ can be derived from the 
determinant of the vierbeins $h$ as follows.  We have
\begin{equation}
h^{\mu}{}_i h^i{}_{\nu} = \delta ^{\mu} {}_{\nu} \,,\,\, h^{\mu} {}_{i} h^{j} {}_{\mu} = \delta _i {}^j\,.
\end{equation}
The relation between  metric and vierbiens  is given by 
\begin{equation}
g_{\mu\nu} = \eta _{ab} h^a {}_{\mu} h^{b} {}_{\nu}\,,
\end{equation}
where $\eta _{ab}$ is the flat Minkowski metric. 
Finally, it is   $|h|\equiv$ det$\left(h^i_\mu\right)=\sqrt{-g}$.  More details  on how the two formalisms are related can be found in \cite{fRT}.

The torsion scalar  is given by the contraction
\begin{equation}
T= S^{\mu\nu} {}_{\rho} T^{\rho} {}_{\mu\nu}\,
\end{equation}
where 
\begin{eqnarray}
S_{\rho} {}^{\mu\nu} &=& \frac{1}{2}\left(K^{\mu\nu}{}_{\rho} + \delta^{\mu} {}_{\rho} T^{\sigma\nu}{}_{\sigma} -\delta^{\nu}{}_{\rho} T^{\sigma\mu}{}_{\sigma}\right)\,, \\
K ^{\mu\nu} {}_{\rho} &=& - \frac{1}{2} \left(T ^{\mu\nu} {}_{\rho} - T^{\nu\mu} {}_{\rho} - T_{\rho} {}^{\mu\nu}\right) \,,\\
T^{\alpha} {}_{\mu\nu} &=& \Gamma ^{\alpha} {}_{\mu\nu} - \tilde{\Gamma} ^{\alpha} {}_{\mu\nu} \,,
\end{eqnarray}
are respectively the superpotential, the contorsion tensor, the torsion tensor and $\tilde{\Gamma} ^{\alpha}{}_{\mu\nu}$ is the Weitzenb\"ock connection. 

Imposing the teleparallelism condition, 
the torsion scalar can be expressed as the sum of the Ricci scalar plus a total derivative term, i.e.
\begin{eqnarray}
 hT = - h\bar{R} + 2 \left(hT_{\nu}{}^{\nu\mu}\right)_{,\mu}\;\; 
\Rightarrow  \;\; T  = - \bar{R} + 2 T_{\nu}{}^{\nu\mu}{}_{,\mu} \,,
\end{eqnarray}
where $\bar{R}$ here is the Ricci scalar corresponding to the Levi-Civita connection and $h$, as above, is the determinant of the metric. Following \cite{Kofinas:2014owa}, 
the teleparallel equivalent of the Gauss-Bonnet topological invariant can be obtained as ,
\begin{equation}
h \mathcal{G} = h T_{\mathcal{G}} + \text{total derivative} \,,
\end{equation}
where the Gauss-Bonnet invariant,  in terms of curvature, is 
\begin{equation}
\mathcal{G} = R^2 -4 R_{\mu\nu}R^{\mu\nu}+ R_{\mu\nu\rho\sigma} R^{\mu\nu\rho \sigma}\,,
\end{equation}
and 
 the teleparallel $T_{\mathcal{G}}$ invariant is given by
\begin{eqnarray}
T_{\mathcal{G}} &=& ( K^{\alpha _1} {}_{ea}K^{e \alpha _2}{}_{b}K^{\alpha _3} {}_{fc}K^{f\alpha _4}{}_{d} - \nonumber \\
&& - 2 K^{\alpha _1 \alpha _2}{}_{a}K^{\alpha _3} {}_{eb} K^{e} {}_{fc}K^{f\alpha _4}{}_{d} + \nonumber \\
&& + 2 K^{\alpha _1 \alpha _2}{}_{a}K^{\alpha _3}{}_{eb}K^{e\alpha _4}{}_{f}K^{f}{}_{cd} + \nonumber \\
&& + 2 K^{\alpha _1 \alpha _2}{}_{a} K^{\alpha _3}{}_{eb} K^{e\alpha _4}{}_{c,d} ) \delta ^a {}_{\alpha _1} {}^b {}_{\alpha _2} {}^c {}_{\alpha _3} {}^d {}_{\alpha _4}\,.
\end{eqnarray}
In a four dimensional spacetime, the term $T_{\mathcal{G}}$ is a topological invariant, constructed out of torsion and contorsion tensor\footnote{See  Section 3 of \cite{Kofinas:2014owa} for the detailed derivation and discussion.}.  In order to simplify the notation,  we will identify $T_{\mathcal{G}}$ with ${\mathcal{G}}$ from now on.

The field equations from the action \eqref{action1} are then
\begin{align}\label{field}
  & 2{f_T}{\partial _\nu }\left( {h{h^\rho }_\kappa {S_\rho }^{\mu \nu }} \right) - 2h{f_T}{h^\gamma }_\kappa {S^{\rho \beta \mu }}{T_{\rho \beta \gamma }} \nonumber \\
  & + 2h{h^\rho }_\kappa {S_\rho }^{\mu \nu }{\partial _\nu }{f_T} + 4h{h_\kappa }^\nu R{R_{\mu \nu }}{f_{\mathcal{G}}} - \frac{1}{2}fh{h^\mu }_\kappa \nonumber \\
   &+ 4h{h_\kappa }^\nu \left( {{g_{\mu \nu }}\square  - {\nabla _\mu }{\nabla _\nu }} \right)\left( {R{f_{\mathcal{G}}}} \right) + 16h{h_\kappa }^\nu {\nabla ^\lambda }{\nabla _{(\mu }}\left( {{f_{\mathcal{G}}}{R_{\nu )\lambda }}} \right)\nonumber \\
    &- 8h{h_\kappa }^\nu {g_{\mu \nu }}{\nabla ^\alpha }{\nabla ^\beta }\left( {{f_{\mathcal{G}}}{R_{\alpha \beta }}} \right) - 8h{h_\kappa }^\nu \square \left( {{f_{\mathcal{G}}}{R_{\mu \nu }}} \right) \nonumber \\
    &- 16h{h_\kappa }^\nu {f_{\mathcal{G}}}{R_{\nu \alpha }}{R^\alpha }_\mu  + 4h{h_\kappa }^\nu {f_{\mathcal{G}}}{R_\nu }^{\alpha \beta \gamma }{R_{\mu \alpha \beta \gamma }} \nonumber \\
    &+ 8h{h_\kappa }^\nu {\nabla ^{(\rho }}{\nabla ^{\sigma )}}\left( {{f_{\mathcal{G}}}{R_{\mu \nu \rho \sigma }}} \right) = 0 
\end{align}
where $f_A=\partial f/\partial A$ being $A=T,\mathcal{G}$. 

In the discussion below, we will consider the Friedmann-Robertson-Walker  (FRW) cosmology  related to $f(T_{\cal G},T)$, i.e. $f({\cal G},T)$, and we search for Noether symmetries in order to fix the form of the function $f$ and to derive exact cosmological solutions.

\section{Searching for Noether Symmetries}
\label{noether}

Let us consider a   a spatially flat FRW  cosmology  defined by the line element
\begin{equation}
\label{FRW}
ds^2 = -dt^2 + a^2(t) (dx^2 + dy^2 + dz^2)\,,
\end{equation}
from which we can express the teleparallel Gauss-Bonnet term as a function of the scale factor $a(t)$ \cite{fGBnoether}
\begin{equation}\label{GBa1}
T_\mathcal{G}= \mathcal{G}= 24 \left[\frac{\dot{a}^2 (t)\ddot{a}(t)}{a(t)^3}\right]\,.
\end{equation}
As said above,  we can discard the total derivative term (see also \cite{sebastian})
The  torsion scalar is
\begin{equation}\label{Ta}
T = - 6\left[\frac{\dot{a}^2(t)}{a^2(t)}\right]\,.
\end{equation}
We can reduce \eqref{action1} to a canonical point-like action by using the Lagrange multipliers as
\begin{equation}\label{eq1}
\mathcal{A} = \frac{1}{2 \kappa}\int dt \left[a^3 f(\mathcal{G},T)- \lambda_1 \left(\mathcal{G} - \bar{\mathcal{G}}\right)- \lambda_2 \left(T - \bar{T}\right)\right]\,,
\end{equation} 
where $\bar{\mathcal{G}}$ and $\bar{T}$ are the Gauss-Bonnet term and the torsion scalar expressed by  \eqref{GBa1} and \eqref{Ta}. The Lagrange multipliers are given by $\lambda_1 = a^3 \partial _{\mathcal{G}}f =a^3 f_{\mathcal{G}}$ and $\lambda_2 = a^3 \partial _{T}f = a^3 f_T$ and are obtained by varying the action with respect to $\mathcal{G}$ and $T$ respectively. We can rewrite  action \eqref{eq1} as
\begin{equation}
\mathcal{A} =  \int \frac{dt a^3}{2\kappa} \left[f(\mathcal{G},T) - f_{\mathcal{G}} \left(\mathcal{G} - \frac{24 \dot a ^2 \ddot a}{a^3}\right)- f_T\left(T+6 \frac{\dot{a^2}}{a^2}\right) \right] 
\end{equation}
and  discarding total derivative terms,   the final Lagrangian is
\begin{equation}\label{PointLagra}
\mathcal{L} = a^3 \left( f - \mathcal{G} f_{\mathcal{G}} - T f_T\right) - 8 \dot a^3 \left(\dot{\mathcal{G}} f_{\mathcal{G}\mathcal{G}}+\dot{T}f_{\mathcal{G}T}\right)- 6 f_T a \dot{a}^2\,,
\end{equation}
This is a point-like, canonical Lagrangian whose configuration space is ${\mathbb {Q}}= \{a,\mathcal{G},T\}$ and  tangent space is ${\mathbb {TQ}} = \{a, \dot{a},\mathcal{G},$ $\dot{\mathcal{G}},T,\dot{T}\}$.
The Euler-Lagrange equations
 for $a,\,\mathcal{G}$ and $T$ are respectively 
\begin{align} 
&{a^2}\left( {f - \mathcal{G}{f_{\mathcal{G}}} - T{f_T}} \right) + 2{f_T}{{\dot a}^2} + 16\dot a\ddot a{{\dot f}_{\mathcal{G}}} + 8{{\dot a}^2}{{\ddot f}_{\mathcal{G}}} +\nonumber \\
&+ 4{{\dot f}_T}a\dot a + 4{f_T}a\ddot a  =0\,, \label{moto11} \\
&\left( {{a^3}\mathcal{G} - 24{{\dot a}^2}\ddot a} \right){f_{\mathcal{G}\mathcal{G}}} + \left( {{a^3}T + 6a{{\dot a}^2}} \right){f_{T\mathcal{G}}} = 0\,, \label{moto22} \\
&\left( {{a^2}T - 6{{\dot a}^2}} \right)a{f_{TT}} - \left( {{a^3}\mathcal{G} - 24{{\dot a}^2}\ddot a} \right){f_{\mathcal{G}T}}  =0\,.
\label{moto33}
\end{align}
As expected, for $f_{\mathcal{G}\mathcal{G}}\neq 0$ and $f_{\mathcal{G}T}\neq 0$, we obtain, from \eqref{moto22} and \eqref{moto33}, the expressions \eqref{GBa1} and \eqref{Ta} for the Gauss-Bonnet term and the torsion scalar. The  energy condition  $E_{\mathcal{L}}=0$, associated with  Lagrangian \eqref{PointLagra}, is
\begin{align}
E_{\cal L}= \frac{\partial {\cal L}}{\partial {\dot a}}{\dot a}+\frac{\partial {\cal L}}{\partial {\dot T}}{\dot T}+\frac{\partial {\cal L}}{\partial {\dot {\cal G}}} {\dot {\cal G}}-{\cal L} &=0\nonumber
\label{energy3}
\end{align}
corresponding to the 00-Einstein equation 
\begin{equation}
24 \dot{a} ^3 {\dot f}_{\mathcal{G}} + 6 f_T a \dot{a} ^2  + a^3 \left( f- \mathcal{G} f_{\mathcal{G}} - T f_T\right) = 0 \,.
\label{energy1}
\end{equation}
Alternatively, the system \eqref{moto11}-\eqref{energy1} can be derived from the field equations \eqref{field}.

Let us now use the Noether Symmetry Approach \cite{cimento} to find possible symmetries for the dynamical system given by  Lagrangian \eqref{PointLagra}.

In general,  a Lagrangian admits a Noether symmetry if its Lie derivative, along a vector field $X$, vanishes\footnote{There exists a symmetry even if the Lagrangian changes by a total derivative term, but we will discuss the simplest case.}
\begin{equation} \label{XL}
L_{X}\mathcal{L} = 0 \Rightarrow X\mathcal{L} = 0 \,.
\end{equation}
Alternatively,  the existence of a symmetry depends on the existence of a vector (a "complete lift"), which is defined on the tangent space of the Lagrangian, i.e.
\begin{equation}\label{cl}
X = \alpha ^i (q)\frac{\partial}{\partial q^i}+\frac{d\alpha^i (q)}{dt}\frac{\partial}{\partial \dot{q}^i}\,,
\end{equation}
being  $q^i$  the configuration variables, $\dot{q}^i$ the generalized velocities and $\alpha^i(q^j)$  the components of the Noether vector. 
In our  case,  the Lagrangian admits three degrees of freedom and then  the symmetry generator \eqref{cl} reads
\begin{equation}
X=\alpha \frac{\partial}{\partial a}+ \beta\frac{\partial}{\partial {\cal G}}+\gamma \frac{\partial}{\partial T}+{\dot \alpha} \frac{\partial}{\partial  \dot a}+  {\dot \beta}\frac{\partial}{\partial\dot {\cal G}}+\dot{\gamma}\frac{\partial}{\partial \dot{T}}\,.
\end{equation}
The system derived from Eq. \eqref{XL} consists of 10 partial differential equations (see \cite{cimento} for details), for $\alpha,$ $\beta,$ $\gamma$ and $f(\mathcal{G},T)$. It is  overdetermined and, if solved, it allows us to determine the components of the Noether vector and the form of $f(\mathcal{G},T)$. It is 

\begin{align}
\label{sys1} &\partial _a \beta f_{\mathcal{G}\mathcal{G}} + \partial _a \gamma f_{\mathcal{G}T} = 0 \,\\ 
\label{sys2} &\beta f_{\mathcal{G}\mathcal{G}\mathcal{G}} + \gamma f_{\mathcal{G}\mathcal{G}T}+ 3 \partial _a \alpha f_{\mathcal{G}\mathcal{G}}+\partial _{\mathcal{G}} \beta f_{\mathcal{G}\mathcal{G}} + \partial _{\mathcal{G}}\gamma f_{\mathcal{G}T} = 0 \,, \\
\label{sys3} &\beta f_{\mathcal{G}T\mathcal{G}} + \gamma f_{\mathcal{G}TT} + 3 \partial _a \alpha f_{\mathcal{G}T} + \partial _T \beta f_{\mathcal{G}\mathcal{G}} + \partial _T \gamma f_{\mathcal{G}T} = 0\,, \\
\label{sys4} &\alpha f_T + \beta f_{T\mathcal{G}}a + \gamma f_{TT} a + 2 f_T a \partial _a \alpha = 0\,, \\
\label{sys5} &a f_T \partial _{\mathcal{G}}\alpha = 0\,,  \\
\label{sys6} &a f_T \partial _{T}\alpha = 0\,, \\
\label{sys7} &f_{\mathcal{G}\mathcal{G}} \partial _{\mathcal{G}}\alpha = 0\,, \\
\label{sys8} &f_{\mathcal{G}T}\partial _T \alpha = 0 \,,\\
\label{sys9}
&\partial _{\mathcal{G}} \alpha f_{\mathcal{G}T} + \partial _T \alpha f_{\mathcal{G}\mathcal{G}} = 0 \,,\\ 
\label{sys10} &3 \alpha \left(f- \mathcal{G}f_{\mathcal{G}}-Tf_T\right) - a \beta \left(\mathcal{G} f_{\mathcal{G}\mathcal{G}}+T f_{T\mathcal{G}} \right) \nonumber \\
&- a \gamma \left(\mathcal{G} f_{\mathcal{G}T}+T f_{TT} \right) = 0 \,.
\end{align}
Clearly, being a system of partial differential equations,  a theorem of existence and unicity for the solutions does not hold.  However, if only one of the functions $\alpha,\beta, \gamma$ is different from zero, a Noether symmetry exists. Below, we will show that the existence of the symmetry selects  the form of the function $f({\cal G},T)$ and allows to get  exact solutions for the dynamical  system \eqref{moto11}-\eqref{energy1}.

\section{Selecting the  form of $f(\mathcal{G},T)$ by symmetries}
\label{selection}

In order to solve   the above system, we have to do some  assumptions. There are two ways to look for solutions: the first, is to assume  specific families of $f(\mathcal{G},T)$ and derive  symmetries accordingly, i.e. find out the components of the symmetry vector.  The second approach consists in  imposing  a specific form for the symmetry  vector  and then   finding  the form of $f(\mathcal{G},T)$. However, in the second case, the chosen functions $\alpha, \beta, \gamma $ must be solution of the system \eqref{sys1}-\eqref{sys10}. To obtain physically reliable models,  the first route can be more convenient.  In this preliminary paper, we will adopt this  strategy to find out solutions choosing classes of $f(\mathcal{G},T)$ function. 

\subsection{The case: $f(\mathcal{G},T)= g_0 \mathcal{G}^k + t_0 T^m$.}
We substitute this form of $f(\mathcal{G},T)$ in the system \eqref{sys1}-\eqref{sys10} and  obtain that for $k\neq 1$ and arbitrary $m$, the only possible Noether vector is the trivial one, $X=(0,0,0)$, which means that there is no symmetry. However, for $k = 1$ and arbitrary $m$, i.e. $f(\mathcal{G},T) = g_0 \mathcal{G}+t_0 T^m$, the vector assume the non-trivial form
\begin{equation}
X\equiv \left\{\alpha _0 a^{1-\frac{3}{2 m}},\,\beta(a,\mathcal{G},T),\,-\frac{3 \alpha _0 T a^{-\frac{3}{2 m}}}{m}\right\}\,,
\end{equation}
with $\alpha _0$ being an arbitrary integration constant and any non singular $\beta$. This means that this theory admits a symmetry with the conserved quantity being
\begin{equation}
\Sigma_0=-12 \alpha _0 m t_0\left( \frac{\dot{a}}{a^{\frac{3}{2 m}-2}} \right)T^{m-1}\,,
\end{equation}
which coincides with the case $f(T)=t_0 T^m$ and then the contribution of the Gauss-Bonnet invariant is trivial\footnote{See Eqs. (455)-(457) in the review paper \cite{RepProgPhys} and the discussion in \cite{greci}.}. This is expected since, in a  4-dimensional manifold,  the  linear  Gauss-Bonnet term  is vanishing in the action and  thus this model is not different from   $f(T)$ gravity.

\subsection{The case: $f(\mathcal{G},T) = f_0 \mathcal{G}^k T^m$.}
In this case, the system \eqref{sys1}-\eqref{sys10} becomes slightly more complicated. As previously, we have two possible choices of the powers $k,m$. If $m\neq 1-k$,  $f(\mathcal{G},T)$ reduces to pure $f(T)$, i.e. we have to set $k=0$ and therefore we have the same symmetries as before. Nevertheless, if $m=1-k$, the model becomes $f(\mathcal{G},T)= f_0 \mathcal{G}^k T^{1-k}$ and it admits a Noether symmetry denoted by the vector 
\begin{equation}
X=(0,\beta(a,\mathcal{G},T),\frac{T}{\mathcal{G}}\beta(a,\mathcal{G},T))\,,
\end{equation}
where $\beta$ is a non-singular function. It is interesting to point out the analogy with the curvature case, where the Noether Symmetry Approach selects the form $f(\mathcal{G},R)= f_0 \mathcal{G}^{1-k}R^{k}$ as discussed in \cite{fGBnoether}. In some sense,  symmetries preserve the structure of gravitational theories independently of the teleparallel or metric formulation\footnote{Clearly also the case 
$f(\mathcal{G},T)= f_0 \mathcal{G}^{1-k}T^{k}$ gives a symmetry.}.

\section{Cosmological solutions}
\label{solutions}
Starting from  the model $f(\mathcal{G},T)= f_0 \mathcal{G}^{k}T^{1-k}$, let us find out  cosmological solutions for any values of $k$. 
The Lagrangian \eqref{PointLagra} assumes the form 
\begin{equation}
\mathcal{L}=f_0 (k-1) \dot{a}^2 \mathcal{G}^{k-2} T^{-k} \left[4 k \dot{a} \left(\mathcal{G} \dot{T} - T \dot{\mathcal{G}}\right)+3 a \mathcal{G}^2\right]\,.
\end{equation}
and the Euler-Lagrange equation for $a(t)$ and the energy equation become
\begin{eqnarray}\label{geneq1}
&& 2 k G^2 a' \left(4 T a'' T'+a' \left(2 T T''-2 k T'^2\right)+a T G'\right) +\nonumber \\
&& +4 k G T a' \left(a' \left(2 (k-1) G' T'-T G''\right)-2 T a'' G'\right)+ \nonumber \\ 
&& +G^3 \left(T \left(2 a a''+a'^2\right)-2 k a a' T'\right) -4 (k-2) k T^2 a'^2 G'^2 =0\,,\nonumber \\
\\\label{geneq2}
&& 4 k a' \left(G T'-T G'\right)+a G^2 =0 \,,
\end{eqnarray}
while the other two, i.e. for $\mathcal{G}$ and $T$ give the Lagrange multipliers \eqref{GBa1} and \eqref{Ta}. If we substitute the constraints \eqref{GBa1},\eqref{Ta} into eq.\eqref{geneq1}, \eqref{geneq2} we get
\begin{eqnarray}
2 a^2 \ddot{a}^4+k^2 \dot{a}^4 \ddot{a}^2-2 (k-1) k a \dddot{a} \dot{a}^3 \ddot{a}+4 k a^2 \dddot{a} \dot{a} \ddot{a}^2 +\nonumber \\
+a \dot{a}^2 \left((k-2) k a \dddot{a}^2+(1-5 k) \ddot{a}^3+k a \ddddot{a} \ddot{a}\right) &=& 0 \,,\\
a \ddot{a}^2 + k a \dddot{a} \dot{a} - k \dot{a}^2 \ddot{a} &=& 0\,. 
\end{eqnarray}
These general (for arbitrary $k\neq 1$) equations admit power law solutions for the scale factor of the form 
\begin{equation}
 a(t) = a_0 t^s,\;\;\;\; \mbox{with}\;\;\;\;s=2k+1\,.
   \end{equation}
 It is easy to verify that the Gauss-Bonnet term and the torsion scalar behave asymptotically as $\mathcal{G} \sim 1/t^4$ and $T \sim 1/t^2,$ for any $k$. 
 
 From these considerations, it is easy to realize that any Friedmann-like, power law solution can be  achieved according to the value of $k$. 
 For example,  a dust solution is recovered for 
 \begin{equation}
 a(t) = a_0 t^{2/3},\;\;\;\; \mbox{with}\;\;\;\;k=-\frac{1}{6}\,;
   \end{equation}
 a  radiation  solution  is for 
 \begin{equation}
 a(t) = a_0 t^{1/2},\;\;\;\; \mbox{with}\;\;\;\;k=-\frac{1}{4}\,;
   \end{equation}
and a stiff matter one is for
 \begin{equation}
 a(t) = a_0 t^{1/3},\;\;\;\; \mbox{with}\;\;\;\;k=-\frac{1}{3}\,.
   \end{equation} 
Power-law inflationary solutions are achieved, in general,  for $s\geq1$ and then $k\geq 0$.   
\section{ Conclusions}
\label{conclusion}  
In this paper, we discussed a theory of gravity where the 
interaction  Lagrangian consists of a generic function $f(T_{\mathcal{G}},T)$ of the teleparallel  Gauss-Bonnet topological invariant, $T_{\mathcal{G}}$, and the torsion scalar $T$. The physical reason for this approach is related to the fact that we want to study a theory where the full budget of torsional degrees of freedom are considered. Furthermore, it is easy to show that, from a dynamical point of view, the Gauss-Bonnet  invariant, derived from curvature, $\cal G$, and the Gauss-Bonnet invariant, derived from torsion, $T_{\mathcal{G}}$, are equivalent and then we can consider a  $f({\mathcal{G}},T)$ theory. 

After these considerations, we searched for Noether  symmetries in the cosmology derived from this model. We showed that  specific forms of $f(\mathcal{G},T)$ admit  symmetries and allow the reduction of the dynamical system. 

In particular, the class $f(\mathcal{G},T)= f_0 \mathcal{G}^kT^{1-k}$ results particularly interesting and, depending on the value of $k$,  it is possible to achieve all the behaviors of standard  cosmology as particular solutions.

Clearly, other cases can be considered and a systematic approach to find out other solutions can be pursued. This will be the argument of a forthcoming paper where a general  cosmological analysis will be developed. 


\section*{Acknowledgements}
The authors acknowledge the COST Action CA15117 (CANTATA) and INFN Sez. di Napoli (Iniziative Specifiche QGSKY and TEONGRAV). M. D. L. is supported by ERC Synergy Grant "BlackHoleCam" Imaging the Event Horizon of Black Holes awarded by the ERC in 2013 (Grant No. 610058). K. F. D. would like to thank the group of Relativistic Astrophysics at the Goethe University (Frankfurt) for the hospitality during the preparation of this paper.

\end{document}